\documentclass[journal,letterpaper]{IEEEtran}
\usepackage{amsmath,amsfonts,amssymb}
\usepackage{algorithmic}
\usepackage{array}
\usepackage{mathtools}
\usepackage{color}
\usepackage{commath}
\usepackage[caption=false,font=footnotesize,labelfont=sf,textfont=sf]{subfig}%normalsize
\usepackage{textcomp}
\usepackage{stfloats}
\usepackage{url}
\usepackage{verbatim}
\usepackage{graphicx}

\title{An Efficient Secret Communication Scheme for the Bosonic Wiretap Channel}

\author{Esther H\"anggi, Iy\'an M\'endez Veiga, and Ligong Wang
\thanks{\scriptsize
Esther H\"anggi is with the Lucerne University of Applied Sciences and Arts, Lucerne School of Computer Science and
Information Technology, 6343 Rotkreuz, Switzerland (e-mail: esther.haenggi@hslu.ch). 
Iy\'an M\'endez Veiga is with the Lucerne University of Applied Sciences and Arts, Lucerne School of Computer Science and
Information Technology, 6343 Rotkreuz, Switzerland and the 
Institute for Theoretical Physics, ETH Zurich, 8093 Zurich, Switzerland
(e-mail: iyan.mendezveiga@hslu.ch). 
Ligong Wang is with the Department of Information Technology and Electrical Engineering, ETH Zurich, 8092 Zurich, Switzerland; this work was conducted when he was with the Lucerne University of Applied Sciences and Arts (e-mail: ligwang@isi.ee.ethz.ch). (All authors are co-first authors.)

This work was supported by the Swiss National Science Foundation Practice-to-Science Grant No 199084.}}

\newcommand{\Ein}{\mathcal{E}}
\renewcommand{\set}{\mathcal}

\makeatletter
\def\@IEEEpubidpullup{3\baselineskip}
\makeatother

\begin{document}
\IEEEoverridecommandlockouts
\IEEEpubid{
\begin{minipage}{\textwidth}
\centering
1558-2558~\copyright~2025 IEEE. All rights reserved, including rights for text and data mining, and training of artificial intelligence\\and similar technologies. Personal use is permitted, but republication/redistribution requires IEEE permission.\\See https://www.ieee.org/publications/rights/index.html for more information.
\end{minipage}
}
\maketitle

\begin{abstract}
We propose a new secret communication scheme over the bosonic wiretap channel. It uses readily  available hardware such as lasers and direct photodetectors. 
The scheme is based on randomness extractors, pulse-position modulation, and Reed-Solomon codes and is therefore computationally efficient. 
It is secure against an eavesdropper performing coherent joint measurements on the quantum states it observes. In the low-photon-flow limit, the scheme is asymptotically optimal and achieves  the same dominant term as 
the secrecy capacity of the same channel.

\end{abstract}


\vspace{-1.5em}
\section{Introduction}
A wiretap channel \cite{wyner75,csiszarkorner78} has one input node, the \emph{sender}, and two output nodes, the \emph{intended receiver} (or simply \emph{receiver}) and the \emph{eavesdropper}. Following the cryptography literature, we shall call them Alice, Bob, and Eve, respectively. Alice wishes to send information reliably to Bob. At the same time, by exploiting the noisy nature of the channel to Eve, this information shall be concealed from Eve. The \emph{secrecy capacity} of the wiretap channel is the largest attainable communication rate with the probability of a decoding error by Bob tending to zero as the number of channel uses grows large, while Eve is kept ``almost completely ignorant'' of the transmitted information.

Like many other results in Information Theory, the secrecy capacity of the wiretap channel was initially derived using probabilistic methods. Later works proposed structured, computationally efficient communication schemes. Among them, the protocol proposed in the series of works \cite{tessaroarxiv,bellare2012cryptographic,semantic} uses \emph{randomness extractors}. These schemes are of polynomial complexity (as opposed to the exponential complexity of the random coding based schemes from \cite{wyner75,csiszarkorner78}). They also reveal interesting connections between information-theoretic and cryptographic approaches to secret communication.

The wiretap channel model has been extended from classical to quantum settings \cite{cai2004quantum,devetak}. Here, we are interested in one specific quantum model, the \emph{pure-loss bosonic wiretap channel}~\cite{smith}. The bosonic channel is often used to model quantum optical communication \cite{shapiro} and therefore of particular practical relevance. The secrecy capacity of the bosonic wiretap channel was established by \cite{smith} and \cite{wolf}. Achieving secrecy capacity generally requires the usage of both advanced hardware (e.g. single photon emitters or joint measurements) and complex algorithms (e.g. random coding). 

We propose a new explicit scheme for secret communication over such channels. For hardware, it only requires lasers and direct detection without photon number resolution. Algorithmically, the scheme uses extractors like in~\cite{tessaroarxiv,bellare2012cryptographic,semantic},  and combines them with pulse-position modulation (PPM) \cite{wangwornell,Kochman_2014} and Reed-Solomon codes. Overall, the scheme is of polynomial complexity. It is hence both physically feasible and computationally efficient.

As we shall see, the proposed scheme achieves the \emph{asymptotic capacity} of said channel in the regime where the number of sent (or received) photons per channel use approaches zero. 

\section{Setup and Background}
\IEEEpubidadjcol
In the pure-loss bosonic wiretap channel, Alice sends a single-mode optical (bosonic) state to Bob through a beam-splitter of transmissivity $\eta\in(0.5,1)$. The remaining optical state that does not reach Bob reaches Eve. In the Heisenberg picture, the channel is characterized as
\begin{subequations}\label{eq:channel}
\begin{IEEEeqnarray}{rCl}
  \hat{b} & = & \sqrt{\eta}\,\hat{a} + \sqrt{1-\eta}\,\hat{v} \label{eq:beamsplitter1}\\
  \hat{e} & = & \sqrt{1-\eta}\,\hat{a} - \sqrt{\eta}\,\hat{v}
\end{IEEEeqnarray}
\end{subequations}
where $\hat{a}$, $\hat{b}$, $\hat{e}$,  and $\hat{v}$ respectively denote the annihilation operators on the Hilbert spaces of Alice, Bob, Eve, and the environment, the last of which being in its vacuum state. Note that, as usual for wiretap channels, Eve is assumed to be passive and cannot influence the channel. 

The channel \eqref{eq:channel} can model all possible types of photon losses, such as path-loss and missed detections, and assumes the worst-case scenario where all photons that do not reach Bob are available to Eve. It does not, however, take into account noise from the environment or the devices.

We impose an average-photon-number constraint on Alice's input state in the form of
\begin{equation}\label{eq:ave}
\langle \hat{a}^\dag \hat{a} \rangle \le \Ein,
\end{equation}
which means that Alice can send, on average, at most $\Ein$ photons in each channel use.

We consider a memoryless setting where the channel can be used many times, and its actions on Alice's input states are independent over time. The power constraint \eqref{eq:ave} is then averaged over the total number of times the channel is used (and also over the message, which is chosen uniformly at random).

The \emph{secrecy capacity} of the channel is the largest rate at which Alice can send information to Bob reliably---meaning that the probability for Bob to decode the message incorrectly will tend to zero as the total number of channel uses grows large---while keeping Eve almost completely ignorant of the transmitted information, i.e., the message is almost uniformly random given the quantum state Eve holds. The secrecy capacity of the channel at hand under constraint \eqref{eq:ave} is given by \cite{smith,wolf}
\begin{IEEEeqnarray}{rCl}
C_\mathrm{s} & = 
& (1+\eta \Ein)\ln (1+\eta \Ein) - (\eta\Ein) \ln (\eta\Ein) \nonumber\\* && {} - \big(1+(1-\eta)\Ein\big) \ln \big(1+(1-\eta)\Ein\big) \nonumber\\*
& & {} + \big((1-\eta)\Ein\big) \ln \big((1-\eta)\Ein\big). \label{eq:capacity}
\end{IEEEeqnarray}
Here and throughout this work we use natural logarithms and information is measured in \emph{nats}. We shall focus on  the regime where $\Ein$ is close to zero. The secrecy capacity \eqref{eq:capacity} then becomes approximately
\begin{equation}\label{eq:capapprox}
C_\mathrm{s} \approx (2\eta -1) \Ein \ln\frac{1}{\Ein}.
\end{equation}

Most schemes achieving secrecy capacity require Alice to send number states (Fock states), or Bob to measure a large number of channel outputs jointly. Both of these are considered difficult to implement in practical scenarios. In contrast, the scheme we propose here can be realized using \emph{coherent states} as Alice's inputs, and \emph{direct detection without photon number resolution} as Bob's measurement, similar to~\cite{wangwornell,Kochman_2014}. 

A coherent state $|\alpha\rangle$, $\alpha\in\mathbb{C}$, can be written in the number-state basis as
\begin{equation}\label{eq:coherent_state}
|\alpha\rangle = e^{-|\alpha|^2/2} \sum_{n=0}^\infty \frac{\alpha^n}{\sqrt{n!}} |n\rangle.
\end{equation}
It describes the optical state emitted by a laser. 
When sent through the channel \eqref{eq:channel}, the state reaching Bob is the coherent state $|\sqrt{\eta}\,\alpha\rangle$, and the state reaching Eve is $|\sqrt{1-\eta}\, \alpha\rangle$. When Bob applies his detector on $|\sqrt{\eta}\,\alpha\rangle$, the output is $0$ (meaning no photon is detected) with probability $e^{-\eta |\alpha|^2}$ and $1$ (meaning one or more photons are detected) with probability $1-e^{-\eta |\alpha|^2}$.

Our goal is to asymptotically attain the secret communication rate \eqref{eq:capapprox} using the above-mentioned transmitter and detector together with computationally efficient encoding and decoding algorithms.

\section{The Scheme}

\begin{figure*}[t]
    \centering
    \includegraphics[width=\textwidth]{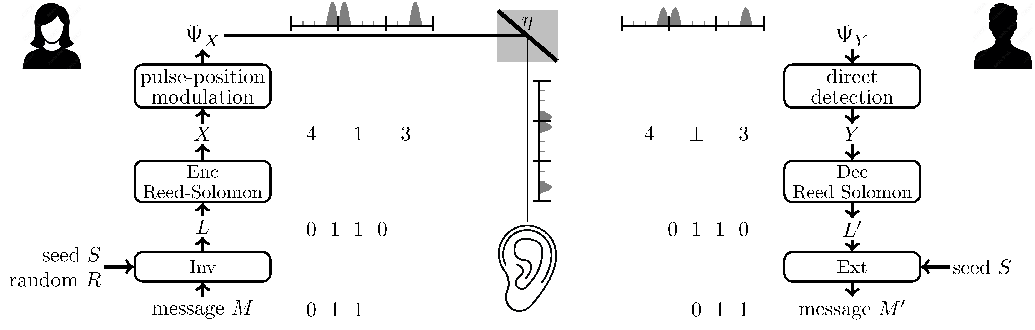}
    \caption{The proposed scheme. Alice applies an inverter of an extractor and then a Reed-Solomon code to the message. She then creates the quantum state $\Psi_X$ using pulse-position modulation. Bob receives the quantum state $\Psi_Y$ and obtains the position of the pulses by direct detection. He then decodes the Reed-Solomon code and applies the extractor to obtain the decoded message.}
    \label{fig:scheme}
\end{figure*}

Our scheme uses the following tools.

\medskip

\emph{Extractor and its inverter.} A quantum-safe strong extractor  $\mathrm{Ext}\colon \set{L}\times\set{S}\to\set{M}$ maps a ``weakly random'' source $L$ and a seed $S$ to a random variable $M$ whose distribution over $\set{M}$ is close to uniform and at the same time almost independent of the seed $S$. 
The Quantum Leftover Hash Lemma~\cite{rennerkoenig,qlhl} states that this holds even \emph{given a quantum state $E^n$} (e.g. held by the adversary) as long as the \emph{smooth quantum min-entropy} of $L$ given $E^n$ is sufficiently high. More precisely, the distance $\Delta$ between the distribution of $M$ (and $S$) conditional on $E^n$ and the uniform distribution is bounded by
\begin{equation} \label{eq:qlhl}
\Delta \leq \frac{1}{2}\sqrt{\abs{\mathcal{M}}e^{-H^\epsilon_{\mathrm{min}}(L|E^n)}}+\epsilon
\end{equation}
for all $\epsilon>0$, where $H^\epsilon_\mathrm{min}$ denotes the \emph{smooth quantum min-entropy} \cite{tomamichelphd}.

The distance $\Delta$ serves as a measure of \emph{secrecy} of our scheme (see~\cite{tessaroarxiv} and~\cite{hanggi2024securityadversarialwiretapchannels}).\footnote{We follow \cite{maurer02} to measure secrecy by comparing the system with an ideal one, in which the message is uniform and independent of the adversary's information.} We emphasize that, by considering the full quantum state Eve obtains over all $n$ channel uses, we take into account joint attacks on all rounds and even allow Eve to delay her measurement until she gains additional information at any later stage.\footnote{These fully general attacks are best compared to \emph{coherent} attacks in quantum key distribution---or, given that the channel to Eve is the same for all rounds, \emph{collective} attacks (see, e.g.~\cite{rennerphd}).} This implies that the achieved security is \emph{composable}~\cite{pw,bpw,canetti}.

We will use extractors that are ``invertible'' in the following sense. An inverter of $\mathrm{Ext}$, denoted $\mathrm{Inv}\colon \set{M}\times\set{S}\times\set{R} \to \set{L}$, takes $M$ and $S$ as given above together with a uniformly distributed random variable $R$, and outputs $L$ which, given $M=m$ and $S=s$, is uniform over the pre-images $\{\ell \colon \mathrm{Ext}(\ell ,s)=m\}$. 

Explicit quantum-safe strong extractors that are efficiently computable~\cite{Schnhage1971} are \emph{finite field extractors}~\cite{carterwegman}: the input and seed are considered as an element of the extension field $\text{GF}(2^\ell)$. The extractor outputs the first $\lambda$ bits of the finite field multiplication of the two. 

\medskip

\emph{Reed-Solomon codes.} These are well-known linear error-correcting codes~\cite{reed_solomon} with computationally efficient encoding and decoding algorithms. The alphabet is a finite field, hence the size of the alphabet $b$ is a prime power.  The block length $n$ must be less than the alphabet size; here we take it to be $b-1$. The message length (of the code) is $k<n$, so there are $k$ message symbols each of size $b$, and the \emph{rate} of the code is $k/n$. The \emph{distance} of the Reed-Solomon code is $n-k+1$, and it can correct up to $n-k$ erasures. Such a code is called a $(b,n,k)$ Reed-Solomon code.

The Reed-Solomon code guarantees that the \emph{probability of error} of our scheme is bounded by the probability that Bob obtains more than $n-k$ erasures.

\medskip

\emph{Pulse-position modulation (PPM).} The channel uses are divided into frames of equal lengths $b$, which will be chosen to equal the alphabet size of the Reed-Solomon code. In each frame, there is only one nonzero channel input (i.e. not in the vacuum state), which we call the ``pulse.'' The pulses in all frames are the same optical state. Thus the input over one frame is specified by an integer from $\{1,\ldots,b\}$ corresponding to the position of the pulse. 

On the receiver side, we record the output in one frame by the position of the (unique) pulse or use $\perp$ to indicate that the pulse is lost, so the output alphabet is $\{\perp,1,\ldots,b\}$.

\medskip

We now describe the secret communication scheme. We assume that, at the beginning of the communication, Alice and Bob share a seed $S$ that is uniform over $\set{S}$. If they do not share a seed, Alice can choose one at random and send it to Bob over the channel publicly. As shown in \cite{semantic}, doing so does not compromise the secrecy of the scheme; nor does it affect the asymptotic communication rate.

Alice's encoding consists of the following steps:
\begin{enumerate}
\item\label{item:alice_inv} Use an inverter $\mathrm{Inv}$, the seed $S$, and local randomness $R$ to expand the message $M$ to a longer string $L=\mathrm{Inv}(M,S,R)$.
\item\label{item:alice_enc} Use a $(b,n,k)$ Reed-Solomon code to encode $L$,  $x^n=\mathrm{Enc}(L)$. (Recall that the alphabet size for $L$ equals $b^k$.)
\item\label{item:alice_ppm} Map each symbol of the Reed-Solomon codeword to a PPM frame with a coherent state $|\alpha\rangle$ at position $x_i\in\{1,\ldots,b\}$ and $|0\rangle$ in the other positions. Note that the total number of channel uses is $n\cdot b$.
\end{enumerate}

Accordingly, Bob's decoding procedure is as follows:
\begin{enumerate}
\item\label{item:bob_ppm} Perform direct detection (without photon number resolution) in each channel use. In each frame, Bob can detect photons at most at one position (where Alice sent a pulse); it can happen that Bob does not detect anything at all. If he does detect photons, then he notes the position of detection as the frame output; if not, then he sets the frame output to be $\perp$, meaning ``erasure.'' Denote the length-$n$ output by $y^n$.
\item\label{item:bob_dec} Apply the Reed-Solomon decoder on the $n$ frame outputs to recover $L'=\mathrm{Dec}(y^n)$.
\item\label{item:bob_ext} Apply $\mathrm{Ext}$, i.e., the extractor corresponding to $\mathrm{Inv}$, on $L'$ to recover the message $M'=\mathrm{Ext}(L',S)$.
\end{enumerate}

\section{Choice of Parameters and Asymptotic Analysis}
Several parameters used in the scheme are related by the constraint \eqref{eq:ave}. In this section, we show how to choose the parameters to achieve a high rate. We compute the asymptotic rate in the low photon regime where $\Ein$ is close to $0$ and show that it reaches the secrecy capacity asymptotically, i.e., that it achieves the dominant term in secrecy capacity.

Our choice is guided by the insights provided in \cite{wangwornell} and \cite{Kochman_2014} regarding the optimal $\alpha$ in the regime where $\Ein$ is small. In fact, all ``$\approx$'' in the following mean that both sides will coincide asymptotically when $\Ein\downarrow 0$.

We choose the size of the PPM frame to be
\begin{equation}
b \approx %\left\lfloor
 \frac{1}{\eta \Ein \ln\frac{1}{\eta\Ein}}. %\right\rfloor.
\end{equation}
More precisely, $b$ should be chosen as the largest prime power not exceeding the right-hand side. 

All the permitted input power over the frame is put into the single pulse $|\alpha\rangle$, so
\begin{equation}
\alpha^2 = b\cdot\Ein \approx \left(\eta \ln\frac{1}{\eta\Ein}\right)^{-1}.
\end{equation} 

At the position where Alice sends $|\alpha\rangle$, after the beam-splitter, Bob receives a coherent state $|\sqrt{\eta}\,\alpha\rangle$ with
\begin{equation}
\eta \, \alpha^2 \approx 
\left(\ln\frac{1}{\eta\Ein}\right)^{-1}
\end{equation}
The probability of erasure---the probability that Bob's detector outputs $\perp$ for the frame---is therefore 
\begin{equation}
\mathrm{Pr}(\textrm{erasure}) = e^{-\eta \alpha^2} \approx 
1 - \left(\ln\frac{1}{\eta\Ein}\right)^{-1}.
\end{equation}
Using that the Reed-Solomon code can correct up to $n-k$ erasures, $k$ can be chosen to be 
\begin{equation}
k \approx (b-1) \cdot (1- \mathrm{Pr}(\textrm{erasure}) ) \approx \frac{1}{ \eta \Ein \left(\ln\frac{1}{\eta\Ein}\right)^2}.
\end{equation}
The total amount of information that Alice can send in the longer string $L$ is then
\begin{equation}
\ln|\mathcal{L}| = k \ln b \approx \frac{1}{ \eta \Ein \ln\frac{1}{\eta\Ein}} \approx b.
\end{equation}

We next estimate how much information is leaked to Eve. Of the photons sent by Alice, a proportion of $(1-\eta)$ reaches Eve (as opposed to $\eta$ that reaches Bob). Each photon, being uniformly distributed in a PPM frame, carries $\ln b$ nats of information. So the total number of nats that are leaked to Eve is approximately
\begin{equation}
b(b-1) \cdot \Ein \cdot (1-\eta) \cdot \ln b \approx \frac{1-\eta}{ \eta^2 \Ein \left(\ln\frac{1}{\eta\Ein}\right)} = \frac{1-\eta}{\eta} b.
\end{equation}
This part of $L$ should be added by the inverter in Alice's encoding scheme, and then removed by Bob using the extractor. That means the number of nats that is contained in the original message $M$ can be at most
\begin{equation}
\ln|\mathcal{M}| \approx  b - \frac{1-\eta}{\eta} b = \frac{2\eta-1}{\eta} b.
\end{equation}
The total number of channel uses being equal to $b(b-1)$, the attained secrecy communication rate is then
\begin{equation}\label{eq:rate}
\textrm{rate} \approx \frac{2\eta-1}{\eta b} \approx (2\eta-1) \Ein \ln \frac{1}{\Ein},
\end{equation}
dropping a $\ln \eta$ term because it is dominated by $\ln \Ein$. This is the same as the approximation given in \eqref{eq:capapprox}. This means that, in the regime where $\Ein\downarrow 0$, the scheme is asymptotically optimal.

\section{Finite Block Length Analysis}\label{sec:finite}

In this section, we derive explicit bounds at finite block lengths on the error probability and the security of the scheme. 

\medskip

\emph{Probability of error.} 
Bob will make a decoding error only when there are more than $n-k$ erasures. The error probability is therefore upper-bounded by the regularized incomplete beta function \cite{paris2010}
\begin{equation} \label{eq:Perror}
\mathrm{Pr}(\mathrm{error}) \le I_{q} (n - k + 1, k)
\end{equation}
with 
$q = e^{-\eta\alpha^2} = \mathrm{Pr}(\textrm{erasure})$ of a single pulse.
When $k = \left\lfloor(1-\theta) (1-e^{-\eta \alpha^2}) n \right\rfloor$ for any small positive $\theta$, this can be bounded by $\mathrm{Pr}(\mathrm{error})\leq e^{-2n\theta^2}$~\cite{chernoff,hoeffding} and
decays exponentially with increasing block length.

\medskip

\emph{Secrecy.} 
By~\eqref{eq:qlhl}, secrecy can be bounded by the $\epsilon$-smooth conditional quantum min-entropy $H_\mathrm{min}^{\epsilon}(L|E^n)$.  
To bound this, we use a chain rule from \cite{tomamichelphd}, which states that, for all $\epsilon'<\epsilon/2$,
\begin{equation}\label{eq:chainrule}
H_\mathrm{min}^{\epsilon}(L|E^n) \ge H_\mathrm{min}(L,E^n) - H_\mathrm{max}^{\epsilon'}(E^n) - 2 \ln \frac{2}{(\epsilon-2\epsilon')^2},
\end{equation}
where $H^{\epsilon'}_{\mathrm{max}}$ is the smooth quantum max-entropy.
Let us now consider each of the terms on the right-hand side. 

The random variable $L$ is uniformly distributed and, conditional on $L$, the eavesdropper's state on $E^n$ is a pure state, so 
\begin{equation} \label{eq:HminLE}
H_\mathrm{min}(L,E^n) = H_\mathrm{min}(L) =  k \ln b.
\end{equation}

To bound $H_\mathrm{max}^{\epsilon'}(E^n)$, note that the number of photons in $E^n$ follows a Poisson distribution with expectation $(1-\eta){\alpha}^2 {n}$. For any $s> (1-\eta){\alpha}^2 {n}$, the probability of observing more than $s$ photons is at most 
\begin{equation} \label{eq:photonnumber}
\Pr[\text{photon number}>s] \leq \frac{2 \gamma\bigl( \lfloor s + 1\rfloor, (1-\eta){\alpha}^2 {n} \bigr)}{\lfloor s\rfloor !}\,,
\end{equation}
with $\gamma$ denoting the lower incomplete $\gamma$-function. 
Consider the projector $\Pi$ onto the subspace of $E^n$ with no more than $s$ photons and let $\tau\triangleq \Pi \rho \Pi$ (without normalization) be the projection of $\rho$ onto this subspace. The purified distance~\cite{tomamichelphd} between $\rho$ and $\tau$ is at most the square root of~\eqref{eq:photonnumber}, i.e.,
\begin{equation}
\epsilon' \leq \sqrt{\Pr[\text{photon number}>s]}\,.
\end{equation} 

It remains to bound the (quantum) max-entropy of $\tau$. In $\tau$, there are at most $\lfloor s \rfloor$ photons, distributed over $n\cdot b$ positions. The max-entropy is the logarithm of the dimension of the image of $\Pi$. This is upper-bounded by
\begin{IEEEeqnarray}{rCl}
\nonumber H_\mathrm{max}^{\epsilon'}(E) &\leq& \ln \left(
\sum_{i=1}^{\lfloor s \rfloor} \binom{nb-1+i}{i} \right)\\
\label{eq:hmaxepsprime} &\leq& (nb-1+s)H_b(\frac{s}{nb-1+s})+\ln s.
\end{IEEEeqnarray}

We obtain the secrecy bound for finite block lengths by applying \eqref{eq:chainrule}, \eqref{eq:HminLE}, and \eqref{eq:hmaxepsprime} to \eqref{eq:qlhl}:
\begin{equation}\label{eq:delta_bound}
\Delta \leq \frac{1}{2}\sqrt{\abs{\mathcal{M}}e^{-k\ln b+ (nb-1+s)H_b(\frac{s}{nb-1+s})+\ln s+2 \ln \frac{2}{(\epsilon-2\epsilon')^2}} }+\epsilon
\end{equation}

\emph{Asymptotics revisited.} We can now reexamine the asymptotic rate for $n,b\to\infty$  and  $\mathcal{E}\downarrow 0$  computed in the previous section. To do so, choose $ s = (1+\delta)(1-\eta)\alpha^2 n$ for any small positive $\delta$ and note that in this case $\epsilon' \leq e^{-\frac{1}{2}\left((1-\eta){\alpha}^2 {n} \right)\left((1+\delta)\ln (1+\delta)-\delta\right)}$ by Bennett's inequality~\cite{Bennett01031962}, which decreases exponentially with $n$. Take $\epsilon$ to be any (small) constant so that the term $2\ln \frac{2}{(\epsilon-2\epsilon')^2}\approx 2\ln\frac{2}{\epsilon^2}$. Then $\Delta$  vanishes as long as 
\begin{IEEEeqnarray}{rCl}
\nonumber \textrm{rate} 
&\leq& \frac{
k\ln b- (nb-1+s)H_b\left(\frac{s}{nb-1+s}\right)-\ln s}{bn}\\
\nonumber & \approx & 
\frac{1-e^{-\eta \alpha^2}}{b}\ln b -\frac{b+(1-\eta)\alpha^2}{b}H_b\left(\frac{(1-\eta)\alpha^2}{b+(1-\eta)\alpha^2}\right)\\
\nonumber
&\approx &
\eta \mathcal{E} \ln b - H_b((1-\eta)\mathcal{E}) \\
\nonumber
& \approx &
\eta \mathcal{E} \ln \frac{1}{\eta\mathcal{E}}-(1-\eta)\mathcal{E} \ln \frac{1}{(1-\eta)\mathcal{E}}\\
\nonumber
& \approx &
\eta \mathcal{E} \ln \frac{1}{\mathcal{E}}-(1-\eta)\mathcal{E} \ln \frac{1}{\mathcal{E}}\nonumber\\*
& & {} +
 \mathcal{E} \left(\ln (1-\eta)+\eta \ln \frac{1}{\eta (1-\eta)}\right) \nonumber
\\
& \approx & (2\eta-1) \mathcal{E} \ln \frac{1}{\mathcal{E}}
\end{IEEEeqnarray}
where we used $k\approx (1-e^{-\eta \alpha^2})\cdot n$, $bn-1+s \approx bn+s$ and $s\approx (1-\eta)\alpha^2 n$ and dropped the term $\ln s$ in the first approximation. 
In the second approximation, we used that $\alpha^2=b\mathcal{E}$ and, since $\mathcal{E}$ is small, $e^{-\eta b \mathcal{E}}\approx 1-\eta b \mathcal{E}$, $1+(1-\eta)\mathcal{E}\approx 1$ and $\frac{(1-\eta)\mathcal{E}}{1+(1-\eta)\mathcal{E}}\approx (1-\eta)\mathcal{E}$. We then used $\ln b \approx \ln \frac{1}{\eta \mathcal{E} \ln \frac{1}{\eta \mathcal{E}}}\approx \ln \frac{1}{\eta \mathcal{E}}$ and the definition of the binary entropy function, dropping the term in $\ln \frac{1}{1-(1-\eta)\mathcal{E}}$. Finally, we dropped the terms that are constant in $ \mathcal{E} $. 
With this, both \eqref{eq:Perror} and \eqref{eq:qlhl} tend to zero, and the largest communication rate allowed by these parameters indeed asymptotically coincides with~\eqref{eq:rate}. 

\begin{figure}[ht]
\centering
\includegraphics[width=\columnwidth]{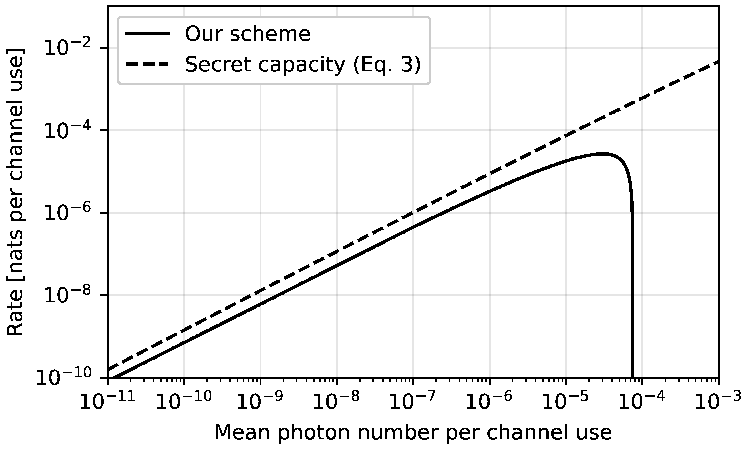}
\caption{Secret capacity~\eqref{eq:capacity} and achievable rates with our scheme for $\eta=0.8$ as functions of $\Ein$. The rates are obtained for $\mathrm{Pr}(\mathrm{error})=10^{-6}$ in~\eqref{eq:Perror} and $\Delta=0.05$ in~\eqref{eq:delta_bound} by optimizing the smoothing parameters $\theta$, $\delta$ and $\epsilon$.}
\label{fig:scheme_rate}
\end{figure}

The secrecy capacity~\eqref{eq:capacity} and achievable rate at finite block lengths are depicted in Figure~\ref{fig:scheme_rate}. As expected, the scheme approaches capacity asymptotically as the mean photon number decreases. The gap vanishes rather slowly; when the mean photon number exceeds a threshold (for the parameters in the plot, it is around $10^{-4}$), our analysis does not guarantee a positive secret communication rate. This is mainly because we are very restrictive on Bob (feasible devices and off-the-shelf decoding algorithms) while assuming a worst-case Eve (collective measurements).

\section{Concluding Remarks}
We proposed a new explicit secret communication scheme over the bosonic wiretap channel that is based on coherent pulses and direct detection; it does not require number state generation, squeezing, collective measurements, etc. It is also computationally efficient. Despite its simplicity, it is asymptotically optimal when the photon flow rate tends to zero. 

The scheme has some limitations. Due to the usage of Reed-Solomon codes combined with PPM, the parameters of the scheme are largely dependent on each other, limiting one's flexibility in choosing transmission power and adapting to the length of the message to be communicated. The parameters can be decoupled if we use other error-correcting codes over PPM; we leave this task for future works. 

Additional directions for future research include
to further narrow the gap between the achievable rate and the secrecy capacity at finite block lengths and therefore realistic transmission power, as well as to extend the scheme to account for errors in the hardware (e.g. ``dark clicks'' of Bob's detector) and thermal noise in the environment. Our method still applies in the presence of such errors, although it would require more advanced decoding algorithms, and the rate analysis would need to be modified accordingly.

\bibliographystyle{IEEEtran}
\bibliography{bosonic.bib}

\end{document}